\documentstyle[manuscript,aps]{revtex}
\begin{document}
\newcommand{\be}{\begin{equation}}
\newcommand{\ee}{\end{equation}}
\newcommand{\bea}{\begin{eqnarray}}
\newcommand{\eea}{\end{eqnarray}}

\newcommand{\bp}{\bar{\phi}}
\newcommand{\ar}{{\rm arsinh}(Ke^{\bp})}
\newcommand{\sq}{\sqrt{K^2 + e^{-2\bp}}}
\newcommand{\arb}{{\rm arsinh}(\bar{K}e^{\bp})}
\newcommand{\sqb}{\sqrt{\bar{K}^2 + e^{-2\bp}}}

\preprint{Freiburg THEP-96/25}

\title{Boundary conditions in quantum string cosmology}
\author{Mariusz P. D\c{a}browski \footnote{E-mail:
             mpdabfz@uoo.univ.szczecin.pl}\\
         {\it Institute of Physics, University of Szczecin, Wielkopolska 15,
          70-451 Szczecin, Poland}
\vspace{5mm}
{\ }\\
Claus Kiefer\footnote{E-mail: kiefer@phyq1.physik.uni-freiburg.de}\\
         {\it Fakult\"at f\"ur Physik, Universit\"at Freiburg, Hermann-Herder-Str. 3,
         D-79104 Freiburg, Germany}}
\date{\today}

\maketitle
\begin{abstract}

We discuss in detail how to consistently impose boundary conditions in 
quantum string cosmology. Since a classical time parameter is absent in 
quantum gravity, such conditions must be imposed with respect to intrinsic 
variables. Constructing wave packets for minisuperspace models from different 
tree-level string effective actions, we explain in particular the meaning 
of a transition between ``pre-big-bang" and ``post-big-bang" branches. This 
leads to a scenario different from previous considerations.

\end{abstract}

\begin{center}
{\it submitted to Physics Letters B}
\end{center}

\newpage

String theory seems to be one of the best current candidates for a theory 
which unifies gravity with other interactions. Since it applies to energies
of the order of the Planck scale, it attracts the interest of cosmologists
who are interested in initial conditions for the universe very close to a
classical singularity. Much interest has been focused recently on 
low-energy effective actions from string theory \cite{fradcall}. Such actions 
contain additional fields in the gravitational sector, in particular dilaton 
and axion fields.

One of the advantages of such an effective theory is the possibility of having a 
superinflationary phase $a(t) \sim (-t)^p$ $(t < 0, p < 0)$, which is driven 
by the kinetic energy of the dilaton, and which is free from the fine-tuning 
problem usually present in potential energy-driven de Sitter or power-law 
inflation. One of the central features of string theory is its symmetry 
with respect to duality transformations \cite{polch}. For simple 
isotropic cosmologies this leads to the scale factor duality 
$(a \to 1/a)$ \cite{vencoll} which, when combined with time reversal 
symmetry, results in new, duality-related solutions. Usually,
one considers one of these
 solutions as describing a superinflationary accelerated 
expansion and the other one as describing a decelerated (presumably radiation 
dominated) expansion. However, the superinflationary phase emerges only for 
negative times $(t < 0)$ and its decelerated duality-related branch is 
separated by a singularity in curvature and string coupling. A desirable 
scenario would be to have a superinflationary phase for negative times 
(the ``pre-big-bang" phase) followed by a standard radiation dominated 
expansion (the ``post-big-bang" phase). However, in view of the appearance 
of the singularity between the two phases, this does not seem to be easily 
achievable. One thus looks for possible mechanisms
 to overcome this ``graceful exit 
problem" in string cosmology \cite{BV}.

It has been proven as a new type of ``no-go" theorem \cite{nogo} that 
there is no way to connect 
classically the duality-related solutions and to overcome the ``graceful 
exit problem" in the simplest models of string cosmology. With respect to 
this result, 
it seems that the classical scenario breaks down and that one needs to take 
quantum effects into account to avoid the singularity. This can be achieved, 
for instance, 
by adopting higher-order $\alpha^{\prime}$ (inverse string tension) 
corrections to the tree-level effective action \cite{fradcall,mel}. Such an 
approach, though preliminary, has been presented recently in \cite{mag}. 
Another possibility is to apply a one-loop superstring effective action, for 
which there exists a large class of nonsingular solutions for
a very broad range 
of the parameters given in \cite{richard}.

Staying on the tree-level sector of the string effective action,
the formalism of canonical quantum gravity has been applied to 
describe a quantum transition form the ``pre-big-bang" phase to the
``post-big-bang" phase through the singularity \cite{GMV,GV}.
More precisely, in the minisuperspace comprising scale factor ($a$)
and dilaton ($\phi$), a solution to the Wheeler-DeWitt equation
was found after imposing boundary conditions in the strong coupling
regime $\phi\to\infty$. This solution was interpreted as describing
a reflection in minisuperspace through the singularity.

Such an interpretation is, however, tight to the presence of an
external time parameter. Being redundant already in the classical
theory due to time-reparametrisation invariance, an external time
parameter is completely absent from quantum gravity (see, for example,
the careful discussion in \cite{zeh1,bar}). A classical time parameter
can only emerge as an approximate notion through some
Born-Oppenheimer type of expansion scheme \cite{semi}.

How, then, can the above ``transition" be consistently dealt with in
quantum cosmology? The choice of boundary conditions as well as
the interpretation of the quantum cosmological wave function
should refer only to {\em intrinsic} variables, i.e. variables
which directly occur in the Wheeler-DeWitt equation. In this respect
 the hyperbolic nature of this equation for such models is
particularly important \cite{zeh1}. The purpose of our paper
is the presentation of a consistent quantum cosmological scenario
along these lines. Instead of referring to an external time,
we shall construct wave packets that represent classical trajectories
in quantum cosmology. This has been successfully applied before
in quantum general relativity \cite{claus1}. Furthermore,
we shall suggest to impose boundary conditions in the region
of small scale factor.

In the following we shall first stick to the simple model where
only a positive cosmological constant is present \cite{GMV,GV}.
The main conceptual issues can be discussed clearly in this context.
We shall then proceed to discuss an example which exhibits
turning points in configuration space.

Before starting with the details of our analysis, we would like
to emphasise that it is not, in general, justified to quantise
an effective action (which itself arises from a fundamental quantum
theory). For example, one would certainly not invoke a 
quantisation of the ``Euler-Heisenberg" effective action of QED.
However, in so far as new fundamental fields arise from the
fundamental theory (such as dilatons and axions), a quantisation
of the effective action could capture some relevant features.
It is with this reservation that we present the following investigation.

We start with the Wheeler-De Witt (WDW) equation from a tree-level 
low-energy string effective action for zero spatial curvature, which 
contains a dilaton potential 
similarly to \cite{GMV,GV}. (Quantum cosmology for string models
was first studied in \cite{BB}. A comparison between Dirac
quantisation and reduced quantisation was made in \cite{CD}.)
 It reads 
\be
\hat{\cal H}\Psi
 \equiv \left[ -\partial_{\bar\phi}^2 + \partial_\beta^2 - \lambda_{s}^2 
V(\beta,\bar{\phi})e^{-2\bar\phi} \right] \Psi(\beta,\bar{\phi}) = 0 \, ,
\ee
where
\bea
\beta & \equiv & \sqrt{3} \ln a \,,\\
\bar\phi & \equiv & \phi - 3 \ln a - \ln \int {d^3x\over\lambda_s^3} \, 
\eea
are redefined variables, 
$V(\beta,\bar{\phi})$ is the dilaton potential, and $\lambda_{s} \equiv 
\sqrt{\alpha^{\prime}}$ is the fundamental string-length parameter
(it is assumed that the volume of three-space is finite).

We first consider the simplest potential which is
just given by a positive cosmological 
constant, $V(\beta,\bar{\phi}) = \Lambda$, and look for a separable solution 
of the form 
\be
\Psi_{k}(\beta,\bar{\phi}) = e^{-ik\beta} \psi_{k}(\bar{\phi})
\ee
for all $k \in {\bf R}$ \cite{GMV,GV}.
 The function $\psi_{k}(\bar{\phi})$ then obeys the effective
equation 
\be
\left( -\partial_{\bar{\phi}}^2 + V_{eff}(\bar{\phi}) \right) 
\psi_k(\bar{\phi}) = 0 \,   ,
\ee
where the effective potential is given by
\be
 V_{eff} = -k^2 - \lambda_{s}^2 \Lambda e^{-2\bar\phi}   .
\ee
Since $V_{eff}$ is always negative, there are no classically 
forbidden regions in the effective sector (as specified by $k$)
 of this simple model and therefore no ``turning points".
(In the full theory, there are no classically forbidden regions,
since the kinetic term is indefinite).
Later we shall discuss other possibilities, where classically forbidden 
regions and turning points exist.

The general solution of (5) is given in terms of Bessel functions \cite{GV},
\be
\psi_k(\bar\phi) = c_1 J_{+ik}(z) + c_2 J_{-ik}(z) \,,
\ee
with $z \equiv \lambda_s \Lambda e^{-\bar\phi}$. In the
strong coupling limit $\bar\phi 
\rightarrow \infty$ $( z \rightarrow 0 )$ one has 
\be
\lim_{z \to 0} J_{\pm ik}(z) e^{-ik\beta} \sim 
e^{ -ik(\beta \pm \bar\phi)}   .
\ee
In order to get a deeper insight into the problem and to discuss the correct 
boundary conditions we include a brief discussion of the classical 
solutions for the string effective action equations 
\cite{vencoll,mel,GMV,GV}. Because of the string duality-symmetry,
 one obtains various
``pre-big-bang" and ``post-big-bang" branches, but we
shall discuss those which 
attracted most interest: the expanding accelerated (or superinflationary) 
``pre-big-bang" branch and the expanding decelerated
 ``post-big-bang" branch, which 
are classically divided by a singularity \cite{GV}. These are given by 

$(+) \;\;\; t < 0$ (``pre-big-bang") 
\bea
\beta & = & \beta_0 - 
 \ln{\tanh{\left(- \frac{\sqrt{\Lambda}t}{2}\right)}}  ,\\
\bar\phi & = & \bar\phi_0 - \ln{\sinh(-\sqrt{\Lambda}t)}   ,
\eea

$(-) \;\;\; t > 0$ (``post-big-bang") 
\bea
\beta & = & \beta_0 + \ln{\tanh{\left(\frac{\sqrt{\Lambda}t}{2}\right)}}  ,\\
\bar\phi & = & \bar\phi_0 - \ln{\sinh(\sqrt{\Lambda}t)}   .
\eea
These branches are related by the duality symmetry including time-reflection 
$\beta(t) \rightarrow - \beta(-t)$, $\bar\phi(t) = \bar\phi(-t)$. 

Since the canonical momentum with respect to $\beta$ is given by 
$\Pi_\beta = -\lambda_{s} e^{-\bar\phi} \dot{\beta} \equiv -k =$ constant, 
one can express ``expansion" by $k = \lambda_{s}
 \sqrt{\Lambda} e^{-\bar\phi_0} > 0$. 
The canonical momentum with respect to $\bar\phi$
 is given by $\Pi_{\bar\phi} = 
\lambda_{s} e^{-\bar\phi} \dot{\bp}$. In the strong coupling regime 
$\bar\phi \rightarrow \infty$ it reads for the cases $(+)$ and $(-)$, 
respectively, 
\be
\Pi_{\bp}^{(\pm)} \stackrel{\bp\to\infty}{\to}
     \pm \lambda_{s} \sqrt{\Lambda} e^{- \bp_0} 
= \pm k   .
\ee
A distinction between ``expanding" and ``contracting" has no intrinsic
meaning, however, since we can arbitrarily change the sign
of $\dot{\beta}$ after re-intro\-du\-cing the lapse-function.
In quantum cosmology, where $t$ is fully absent, this becomes even
more evident, since no reference phase $\exp(-i\omega t)$
is available, with respect to which solutions could be classified
as, e.g., right-moving or left-moving. This is in full analogy
to the situation in ordinary quantum cosmology \cite{zeh1,claus2,zeh2}.

To make the identity of an expanding solution with a contracting
solution explicit, it is more
appropriate to discuss the string scenario in the configuration space formed 
by $(\beta, \bar\phi)$. Eliminating $t$ in (9)--(12), one finds that the 
trajectories in configuration space are given by 
\be
\beta = \beta_0 \pm {\rm arsinh}(Ke^{\bar{\phi}}) = 
\beta_0 \pm \ln{\left[ \left( K + \sqrt{K^2 + e^{-2\bar{\phi}} } \right)
e^{\bar{\phi}}\right]}   ,
\ee
where the plus sign refers to the ``pre-big-bang" branch $(+)$
 and the minus sign
to the ``post-big-bang" branch $(-)$, respectively, and a new constant $K$,  
\be
K \equiv \frac{k}{\lambda_{s}\sqrt{\Lambda}} = \pm e^{-\bar\phi_0}   ,
\ee
has been introduced. (A change of sign of $K$ corresponds to
the change of the branch, the above distinction between $(+)$
and $(-)$ thus holding for $K>0$. This is the case to which we
restrict our analysis without loss of generality.)
Therefore, we still have the two branches in configuration space 
which tend to the same limit $\beta = \beta_0$ in the low-energy regime 
$\bar\phi \rightarrow - \infty$. 

What we called ``pre-big-bang" branch (``post-big-bang" branch)
is now the upper (lower) branch in configuration space,
and the duality transformation transforms $\beta(\bp)-\beta_0
\to \beta_0-\beta(\bp)$. We note that the qualitative
features of the trajectories remain unchanged if we go back
to the original configuration space variables $\beta$ and $\phi$,
where $\phi$ is the original dilaton field, see Eq.~(3). The two branches 
are then given by
the equation (if we write $\bp=\phi-\sqrt{3}(\beta-\beta_0)$)
\[ e^{\phi}=\vert K\vert^{-1}e^{\pm(\beta-\beta_0)\sqrt{3}}
   \sinh(\pm\beta\mp\beta_0). \]

Coming back to the solution (7) of the effective equation (5) and its 
limit (8), one can see that 
\be 
\lim_{\phi \to \infty} \Pi_{\bar\phi} J_{\pm ik}(z) = \mp k J_{\pm ik}  ,
\ee
where $\Pi_{\bar\phi} = -i \partial_{\bar\phi}$. This quantum relation, 
or more precisely, its analogy with the classical relation (13) was the 
key point in \cite{GV} to identify the two 
solutions $J_{\pm ik}$ with the ``pre-big-bang"  
($J_{-ik}$) and the ``post-big-bang" ($J_{+ik}$) solutions, respectively. 


As we have argued before, however, such a distinction has no intrinsic 
meaning. One can only talk about plane waves travelling {\it with respect to}
the ``intrinsic time" $\beta$, distinguishing small $\beta$ and large $\beta$,
 but not {\em the} ``pre-big-bang" and {\em the} ``post-big-bang".

In order to gain further insight, we try to construct {\it wave packets} 
following the classical trajectories in configuration space given by (14). 
For the sake of this purpose it is convenient to study first a WKB 
approximation to (5). Since there are no classically forbidden regions, 
one has for all values of $\bar\phi$,
\be 
\psi_k(\bar\phi) \sim \left( - V_{eff} \right)^{-\frac{1}{4}} 
\left[ \exp{\left(i \int {\sqrt{-V_{eff}}d\bar\phi}\right)} +
C \exp{\left(- i \int {\sqrt{-V_{eff}}d\bar\phi}\right)} \right]  ,
\ee
where $C$ is a
 constant, and ``$\exp(+)$" refers to ``pre-big-bang", while 
``$\exp(-)$" refers to ``post-big-bang".
 The total WKB phase ($\psi_k \sim e^{iS_k}$) 
is then
\be
S_{k}^{(\pm)}(\beta, \bar\phi) = - k \beta \pm s_k(\bar\phi), 
\ee
where
\be
s_k \equiv \int^{\bar\phi} \sqrt{k^2 + \lambda_s^2 \Lambda e^{-2\tilde{\phi}}}
d\tilde{\phi}   .
\ee
This integral can be solved exactly to give
\be
s_k = \lambda_s \sqrt{\Lambda} \left\{K \left[ {\rm arsinh}(Ke^{\bp}) \right]
 - \sqrt{K^2 + e^{-2\bp}} \right\}   ,
\ee
By the principle of constructive interference \cite{claus1}, the classical 
solutions are found through
\be
\frac{\partial S_k^{(\pm)}}{\partial k}
 = - \beta \pm \frac{\partial s_k}{\partial k}
= 0   ,
\ee
leading to Eq.(14) for the classical trajectories in configuration space. 
After rescaling $S_k^{(\pm)} \rightarrow
 S_k^{(\pm)}/\lambda_s \sqrt{\Lambda}$ we have for 
the total WKB phase (18)
\be
S_k^{(\pm)} = - K\beta \pm K \ar \mp \sq   .
\ee
In order to calculate wave packets for the two solutions (22) we take a 
Gaussian concentrated around $\bar{K}>0$, 
\be
A_k = \pi^{-\frac{1}{4}}b^{-\frac{1}{2}} \exp{\left[ -\frac{1}{2b^2} 
\left(K - \bar{K}\right)^2 \right]}
\ee
and consider the superposition
\be
\Psi^{(\pm)}(\beta,\bp) = \int_{-\infty}^{\infty} dK A_k
\frac{e^{iS_k^{(\pm)}}}{(-V_{eff})^{\frac{1}{4}}}   .
\ee
If the width $b$ of the Gaussian is small enough, $A_k$ is concentrated around 
$K \approx \bar{K}$, and therefore the integral (24) can be evaluated by 
expanding $S_k^{(\pm)}$ up to quadratic order in $K - \bar{K}$. Then, 
\be
iS_k^{(\pm)} = -iK\beta \pm iK \arb \mp i\sqb \pm 
i\frac{(K - \bar{K})^2}{2\sqb} + \ldots     .
\ee
Inserting this into (24) and evaluating the resulting Gaussian integral, we 
have (choosing $\beta_0 = 0$ for simplicity) 
\bea
\Psi^{(\pm)}(\beta,\bp) & = & \sqrt{\frac{2}{\pi b}} \frac{1}{B} 
\left(\bar{K}^2 + e^{-2\bp}\right)^{- \frac{1}{4}} 
\exp\left[-i\bar{K} \left(\beta \mp \arb \right)\right. \nonumber\\
 & & \left. \mp i \sqb \hspace{0.1cm}
\right] \times 
\exp{\left[\frac{1}{2B^2}\left(-\beta \pm \arb \right)^2 \right]}   ,
\eea
where 
\be
B^2 = \frac{1}{b^2} \mp \frac{i}{\sqb}   .
\ee
It is obvious that $\vert \Psi^{(\pm)} \vert^2$ is peaked around the classical 
trajectories (14) in configuration space. The absolute square of the width 
$B$ is given by
\be
\vert B \vert^2 =
 \frac{1}{b^2} \sqrt{ 1 + \frac{b^4}{\bar{K}^2 + e^{-2\bp}}}  ,
\ee
so we have a very ``mild spreading" of the packet. 

We consider now packets from exact solutions. The strong coupling limit 
$\bp \rightarrow \infty$ $(z \rightarrow 0)$ was already performed in (8), 
while in the low energy limit 
$\bp \rightarrow - \infty$ $(z \rightarrow \infty)$ we have
\be
J_{\pm ik}(z) = \sqrt{\frac{2}{\pi z}} \cos{\left(z \mp \frac{\pi ik}{2} - 
\frac{\pi}{4} \right)} + \ldots \propto
\frac{1}{2} e^{\pm \left( \frac{\pi k}{2} - \frac{i\pi}{4} \right)}
\left(e^{iz} + ie^{-iz \mp \pi k} \right)   .
\ee

The corresponding wave packets read
\be
\Psi^{(\pm)}(\beta,\bp) = \int_{-\infty}^{\infty} dK A_k
e^{-ik\beta}J_{\mp ik}(z)  .
\ee
Following the discussion of \cite{GV} we note that after taking, for instance,
the ``pre-big-bang" solution $J_{-ik}(z)$ for $\bp \rightarrow -\infty$ 
$(z \to \infty)$, one has a superposition of $(+)$ and $(-)$ solutions 
(cf. Eq.(29)),
\be
J_{-ik}(z) \propto  e^{iz} + ie^{-iz}e^{\pi k} \equiv (-) + (+),
\ee
and therefore the relative probability between $(+)$ and $(-)$ is 
\be
\frac{\mid \Psi_{\bp \to -\infty}^{(-)} \mid^2}
{\mid \Psi_{\bp \to -\infty}^{(+)} \mid^2} 
= e^{-2\pi k}   .
\ee
However, in order to have a sensible wave 
packet, $k$ should be concentrated around $k \gg 1$. This means 
that a ``transition" into the $(-)$ component for $\bp \rightarrow -\infty$ 
could only be interpreted as an extremely unlikely quantum effect
in that region, but 
{\it not} as a transition into
 the other semiclassical component as represented 
by a wave packet. Roughly speaking, the $(-)$-component does not correspond 
to a ``classical" trajectory if $J_{-ik}$ is chosen as the exact solution.

To achieve interference between $(+)$ and $(-)$ wave packets, one must really 
{\it superpose} both packets, 
\be
\Psi = \alpha_1 \Psi^{(+)} + \alpha_2 \Psi^{(-)}, 
\ee
i.e., choose
\be
\Psi^{(\pm)}(\beta,\bp) = \int_{-\infty}^{\infty} dK A_k
e^{-ik\beta}\left[ \alpha_1 J_{-ik}(z) + \alpha_2 J_{+ik}(z) \right]   
\ee
with complex coefficients $\alpha_1$ and $\alpha_2$. 
This happens, for example, if boundary conditions are imposed in the low 
energy limit $\bp \rightarrow - \infty$ instead of the strong coupling 
limit $\bp \rightarrow \infty$, in contrast to \cite{GV}.
(This is also the region where the effective theory
can be trusted.) A superposition 
of $J_{+ik}$ and $J_{-ik}$ 
which corresponds to the (+) solution for $\bp \to -\infty$ (compare (31)) 
is the Hankel function 
\be
H_{ik}^{(2)}(z) \stackrel{z\to\infty}{\sim} \sqrt{\frac{2}{\pi z}}
 e^{-iz - \frac{\pi k}{2} + 
\frac{i\pi}{2}}   .
\ee
Since 
\be
H_{ik}^{(2)}(z) = J_{ik}(z) - i N_{ik}(z) = \left(1 - \coth{k\pi}\right) 
J_{ik} + \frac{J_{-ik}}{\sinh{k\pi}}   ,
\ee
one finds that $H_{ik}^{(2)}(z)$ approaches in the strong coupling 
limit $\bp \rightarrow \infty$ $(z \to 0)$ the following asymptotic 
behaviour: 
\be
H_{ik}^{(2)}(z) \rightarrow \frac{1}{\sinh{k\pi}} \left[ 
\frac{e^{-k\pi}}{\Gamma(1 + ik)}
\left(\frac{z}{2}\right)^{ik} + \frac{1}{\Gamma(1 - ik)}
\left(\frac{z}{2}\right)^{-ik} \right]  .
\ee
The corresponding ``transition factor" from $(+)$ to $(-)$ would then
again be given by $e^{-2\pi k}$, but this time a second semiclassical
component is indeed present. This is a generic feature if boundary
conditions are imposed at $\bp\to-\infty$: since the classical solutions
overlap in this region, one finds in general a superposition of
wave packets for $\bp\to\infty$. 

We want to include here a general discussion of boundary
conditions in quantum string cosmology.
If more than two degrees of freedom are present (which, of course,
is the realistic case), the Wheeler-DeWitt equation is, at least 
for perturbations of Friedmann-type spaces \cite{nico},
hyperbolic with respect to $\beta$. One would thus expect
to impose boundary conditions (Cauchy data) at $\beta=constant$ (or
$a=constant$). 

As long as one considers only minisuperspace degrees of freedom,
the wave packets are just timeless wave tubes. A semiclassical
time parameter, as well as the concept of a {\em direction} of
time can only be defined if a huge number of further degrees of
freedom (``higher multipoles") is present. A semiclassical
time parameter emerges if the ``background wave function"
is in a WKB state \cite{semi}. Technically this is achieved by
a Born-Oppenheimer type of expansion scheme, with the
expansion parameter given by $\lambda_s$ in the present case.
It yields $\partial/\partial t\equiv\nabla S\cdot\nabla$
for background states $\psi_0\approx e^{iS}$. A time direction
then emerges from thermodynamical considerations if one starts
from an uncorrelated state for $\beta\to-\infty$. Such an
``initial condition" is facilitated by the fact that the potential
term in the Wheeler-DeWitt equation vanishes in this limit
(except for the dilaton part). Such an initial state can,
for example, be of the form $\Psi=\psi_0(\beta,\phi)$, independent
of other degrees of freedom (see \cite{CZ} for the case of quantum
general relativity).
 With increasing values of $\beta$,
a correlated state would emerge, since the potential now depends
explicitly on the higher multipoles. This in turn, leads to
decoherence and increasing entropy for the background part
$(\beta,\phi)$ \cite{claus2,deco}.

For higher values of $\beta$ one will then enter the semiclassical
regime. One will then get, for example, a state of the form
\[ \Psi\approx \alpha_1e^{iS^{(+)}(\beta,\phi)}\chi^{(+)}
   (\beta,\phi,\{x_{\lambda}\}) +\alpha_2
  e^{iS^{(-)}(\beta,\phi)}\chi^{(-)}
   (\beta,\phi,\{x_{\lambda}\}), \]
where $\{x_{\lambda}\}$ symbolically denotes all higher multipoles.
It is then a quantitative question whether there will be
also decoherence between these two components, in addition to the
decoherence for each single component. It has been argued that
there are regions in the $(\beta,\phi)$-plane
 (concentrated towards negative dilaton values) where decoherence
is ineffective \cite{LP}. It would nevertheless then be inappropriate
to imagine this as a ``transition" from one semiclassical component
into the other, since the semiclassical approximation breaks
down in such a region, so that no time parameter exists there.
There thus exists no classical causal relationship between
these branches. This makes it very hard to solve the
``graceful exit problem" in this framework, and one has
to envisage alternatives such as in \cite{mag,richard}.

In the last part of our paper we want to discuss briefly
 some other possibilities for the dilaton potential 
$V(\beta,\bp)$ in the WDW equation~(1) in order to gain insight into 
the problem of boundary conditions in other situations. First, we adopt 
(though 
hardly justified by string theory) the negative dilaton potential \cite{GMV}
\be
V(\beta,\bp) = - V_0 e^{4\bp}  \hspace{2.0cm} (V_0 > 0)
\ee
in (1), which allows one to find the separable solution (4) with $\psi_k(\bp)$
obeying the effective equation (5). But now the effective potential is different 
from that of (6) and reads 
\be
V_{eff} = -k^2 + \lambda_s^2 V_0 e^{2\bp}  .
\ee
The potential (39) leads to the existence of
 classically forbidden regions and   
``turning points". The key point of such a model
 is that the ``pre-big-bang" 
and the ``post-big-bang"
 branches are already connected at the classical level.
It is, however, interesting that, in contrast to most situations
in ordinary cosmology, 
it is not the scale factor $a$, but the shifted dilaton $\bp$ which has 
a turning point. 
Another point is that the strong coupling limit $\bp \rightarrow \infty$ 
is classically forbidden. 

The corresponding classical self-dual solution \cite{GMV} is given by 
\bea
\bp & = & -\frac{1}{2} \ln{\left(
\frac{V_0^{\frac{1}{2}}}{L^2} + L^2 V_0^{\frac{1}{2}} t^2 \right)}   
,\\
\beta & = & \beta_0 + L^2 t + \sqrt{1 + L^4 t^2}   ,
\eea
where 
\be
L = \frac{k}{\lambda_s V_0^{{1}/{4}}}   .
\ee
This solution is nonsingular at $t = 0$, and the evolution 
of the scale factor seems to describe a transition between
 the ``pre-big-bang" 
accelerated branch and the ``post-big-bang"
 decelerated branch of singular solutions
like (9)-(12). After eliminating $t$ from (42)-(43) we find for the evolution 
equation in configuration space $(\beta, \bp)$ the equation\footnote{Note
 that there 
is also a self-dual solution which
 connects smoothly the ``pre-big-bang" decelerated 
branch with the ``post-big-bang" accelerated branch
 with the same Eq. (40) and 
the opposite sign before $L^2t$ in (41), which leads to the same evolution 
equation in configuration space (43).}
\be
\beta - \beta_0 = \pm {\rm arcosh} \frac{e^{\bp_0}}{e^{\bp}}   ,
\ee
where we have defined
\be
e^{\bp_0} \equiv \frac{L}{V_0^{{1}/{4}}}   .
\ee

The trajectory (43) cannot be divided into two branches
 which could be naturally interpreted as describing 
``pre-" or ``post-big-bang" branches. In particular, the shifted 
dilaton has a ``turning point" for $\bp = \bp_0$. This equation looks 
similar to the corresponding equation in the case of a
massless scalar field in ordinary cosmology \cite{claus1},
except that the roles of field and scale factor are interchanged.
For this reason we have here a ``turning point" for the
dilaton. As discussed above, $\beta$ plays the role of an
intrinsic time variable. A sensible boundary condition would then
be to have a wave packet in the small $\beta$-region concentrated
around a large negative value for the dilaton.
This would then lead to a wave packet concentrated around
the trajectory depicted in Fig~2. We emphasise again that this
can only be interpreted as representing {\em one} cosmological
solution. This solution can be labelled ``expanding"
only after a condition of low entropy is imposed for
$\beta\to-\infty$ in the sense discussed above.

An interesting model with a turning point in $\beta$ is obtained
by taking into account a positive curvature term in the
effective action \cite{EMW}. We shall not, however, include a
discussion of this model here, since the essential conceptual
features remain unchanged.

\acknowledgments

MPD wishes to thank David Wands for useful discussions. He also thanks the 
DAAD (Deutscher Akademischer Austauschdienst) for financial support during 
his visit to the University of Freiburg.

\end{document}